\documentclass[12pt,psfig,a4]{amsart}

\usepackage{geometry}
\usepackage{amscd}
\usepackage{graphics}
\newtheorem{theorem}{Theorem}[section]
\newtheorem{lemma}[theorem]{Lemma}
\newtheorem{pro}[theorem]{Proposition}

\theoremstyle{definition}

\theoremstyle{remark}

\numberwithin{equation}{section}

\begin{document}
\bibliographystyle{plain}
\pagestyle{plain}
\pagenumbering{arabic}

\title{A quantization procedure based on completely positive maps and Markov operators}

\author{Carlos F. Lardizabal}

\address{Instituto de Matem\'atica - Universidade Federal do Rio Grande do Sul - UFRGS - Av. Bento Gon\c calves 9500 - CEP 91509-900 Porto Alegre, RS, Brazil}
\curraddr{}
\email{cfelipe@mat.ufrgs.br}




\date{}

\dedicatory{}

\begin{abstract}
We describe $\omega$-limit sets of completely positive (CP) maps over finite-dimensional spaces. In such sets and in its corresponding convex hulls, CP maps present isometric behavior and the states contained in it commute with each other. Motivated by these facts, we describe a quantization procedure based on CP maps which are induced by Markov (transfer) operators. Classical dynamics are described by an action over essentially bounded functions. A non-expansive linear map, which depends on a choice of a probability measure, is the centerpiece connecting phenomena over function and matrix spaces.


\end{abstract}

\maketitle

\def\laa{\langle}
\def\raa{\rangle}
\def\bv{\big\vert}
\def\SP{\hspace{1.5cm}}
\def\sp{\hspace{0.2cm}}
\def\qed{\begin{flushright} $\square$ \end{flushright}}
\def\qee{\begin{flushright} $\Diamond$ \end{flushright}}
\def\ov{\overline}

\section{Introduction}

In quantum mechanics there exists a great interest in problems in which the system and the surrounding environment are not isolated. Whenever we have that interactions between a system S and environment E cannot be ignored, we have an {\it open system}. The dynamics associated to the (reduced) system $S$ is irreversible and incorporates noise and dissipative effects due to the presence of $E$. When considering open quantum systems, a  starting point is the associated Master equation \cite{nielsen},
\begin{equation}
\frac{d\rho}{dt}=-\frac{i}{\hbar}[H,\rho]+\sum_j[2L_j\rho L_j^* -\{L_j^*L_j,\rho\}],
\end{equation}
where $\{A,B\}=AB+BA$, $H$ is hermitian (coherent part of the dynamics) and $L_j$ represents the coupling between $S$ and $E$ (Lindblad operators). Also, it is well-known that solutions for Master equations can be written in terms of completely positive maps (CP), $\Phi(\rho)=\sum_i V_i\rho V_i^*,$ where the $V_i$ are linear. In general, CP maps are not unitary and it is of interest to analyze the asymptotic evolution of the associated dynamics. This has been done for finite-dimensional CP maps in \cite{novotny}, in the case of mixed-unitary operators, and \cite{petulante} for bistochastic operators. A related question on CP dynamics is to describe the $\omega$-limit sets, that is, the set of density matrices which are limit of $\Phi^{n_i}(\rho_0)$ for a fixed density $\rho_0$, and for some subsequence $\{n_i\}$. There are well-known results on limit sets coming from the theory of non-expansive maps over Banach spaces, so it is natural to pursue such description on CP maps, since these are non-expansive with respect to the trace distance, $d(\rho,\sigma)=\frac{1}{2}tr|\rho-\sigma|$, see \cite{ruskai}.

\bigskip

Motivated by certain limit results, we consider a classical map which acts on densities (i.e., positive functions on a compact space). Assume such dynamics is described by a linear, positive map. We ask the following question: is there a CP evolution on some matrix space which can be seen as the quantum counterpart of the classical evolution? In this work we describe a quantization procedure based on CP maps. We will require that fixed points of the classical dynamics are taken to fixed points in the quantum setting, and similarly to $\omega$-limit sets. The procedure connects classical and quantum dynamics in the following way:
\begin{equation}\label{cdd}
\begin{CD}L^\infty(E_M;\mu) @>L>> L^\infty(E_M;\mu) @>L>> \cdots \\
@V{\kappa_\mu}VV  @V{\kappa_\mu}VV \\
I_n\otimes M_n @>{\Lambda}>> I_n\otimes M_n @>{\Lambda}>> \cdots
\end{CD}
\end{equation}
The set $L^\infty(E_M;\mu)$ denotes the set of essentially bounded functions with respect the measure $\mu$ over the space $E_M$ of state functionals over $M=M_n(\mathbb{C})$, the order $n$ matrices. The upper line represents the classical evolution and the operator $L$ is the transfer operator. Its dual, acting on probability measures, is the associated Markov operator. The lower part contains $\Lambda$, which is a CP map to be determined. The map $\kappa_\mu$ is defined as follows. Let $\mu\in M_\rho(E_M)$, a probability measure over $E_M$ with barycenter $\rho$. Then $f\in L^\infty(\mu)\mapsto \kappa_\mu(f)\in I_n\otimes M_n(\mathbb{C})$ is defined by:
\begin{equation}
\langle\Omega_\rho,\kappa_\mu(f)\pi_\rho(A)\Omega_\rho\rangle=\int f(\omega')\omega'(A)\; d\mu(\omega')
\end{equation}
Such map is linear and non-expansive \cite{bratteli}. We choose a faithful state $\rho\in M_n$ and set the GNS triplet as $(H_\rho,\pi_\rho,\Omega_\rho)=(\mathbb{C}^n\otimes\mathbb{C}^n, M_n\otimes I_n,|\sqrt{\rho}\raa)$. By this we mean that $\pi_\rho(M_n)=M_n\otimes I_n$, where if $\rho=\sum_i r_i |r_i\raa\laa r_i|\in M_n$, then define
\begin{equation}
|\sqrt{\rho}\raa=\sum_i \sqrt{r_i}|r_i\raa\otimes |r_i\raa\in \mathbb{C}^n\otimes\mathbb{C}^n
\end{equation}
For a certain class of measures (orthogonal), Tomita's theorem \cite{bratteli} states that $\kappa_\mu$ is a $*$-isomorphism into $I_n\otimes M_n$. But we note that the description presented in this work is also of interest even when the measures involved are not orthogonal. Given $L$ a transfer operator, a positive function $f$ and a probability measure $\mu$, we wish to find $\Lambda=\Lambda(\mu, f)$ such that the above diagram commutes. We note that $E_M=E_{M_n}$, the space of state functionals over $\mathbb{C}^n$, can be identified with the space $D(\mathbb{C}^n)$ of density matrices in $M_n(\mathbb{C})$. We prove in section \ref{qua_proc}:

\bigskip

{\bf Theorem \ref{the_theorem}.} Let $L: C(E_{M_n})\to C(E_{M_n})$, $L(g)(\rho)=\sum_{i=1}^k p_i g(F_i(\rho))$, $k\leq n^2$, $p_i\geq 0$.
\begin{enumerate}

\item For a given $f>0$ and for every $\mu=\sum_m\lambda_m\delta_{\omega_m}$, $\lambda_m\geq 0$, $\omega_i$ with support on the diagonal units of $M_n(\mathbb{C})$, there exists a solution for $\Lambda\circ \kappa_\mu(f)=\kappa_\mu(Lf)$, given by $\Lambda(\rho)=\sum_{i=1}^k V_i\rho V_i^*$, $V_i=\textrm{diag}(v_1^i,\dots, v_n^i)$, and
\begin{equation}
|v_j^i|=\sqrt{\frac{p_i\Big(\sum_m\lambda_m f(F_i(\omega_m))\omega_m(E_{jj})\Big)}{\sum_m\lambda_m f(\omega_m)\omega_m(E_{jj})}},\;\;\; i=1,\dots, k,\;\;\; j=1,\dots,n
\end{equation}

\item If $Lf=\sum_i p_i f(F_i)$ with $F_i=U_i\cdot U_i^*$, $U_i$ unitary, there exists $V_i$ which satisfies $\Lambda\circ \kappa_\mu(f)=\kappa_\mu(Lf)$ for every $f\in L^\infty$ if and only if the $\omega_i$ are all fixed points for $\Phi(\rho)=\sum_i p_iF_i(\rho)$.
\end{enumerate}

\bigskip

Among the known connections between Markov operators and quantum systems, we recall that a density matrix $\rho_0$ is a fixed point for a mixed unitary channel $\Lambda$ if and only if $\rho_0$ is the barycenter of a measure which is invariant for the associated Markov operator \cite{lozinski}, \cite{slomcz}. We are interested in further developing such connection, having in mind applications to quantum information theory. In this article we focus on finite-dimensional quantum channels.

\bigskip

The structure of this work is the following. In section \ref{cp} we review basic facts on completely positive maps. In section \ref{omegalims} we review $\omega$-limit sets, results on non-expansive maps, and describe the related results for CP maps which are of our interest. Several of the results in that section are easy adaptations of well-known theorems on non-expansive maps on Banach spaces and the proofs are briefly described in the appendix. In section \ref{meas_ortho} we review some facts on measure theory and representations. In section \ref{tomita} we review the transfer operator, which represents the action on the space of functions. In section \ref{conditions_on} we describe the quantization procedure and conditions for its existence. We conclude in section \ref{conclusion} with some open questions on a classical commutative diagram and its relations to the construction made in this work.

\section{Evolution of classical and quantum dynamics}

\subsection{Completely positive maps}\label{cp}

We make a brief review on completely positive maps \cite{petz}. For simplicity we will often restrict ourselves to the space $M_n(\mathbb{C})$ of order $n$ matrices. Let $\mathcal{H}$ be a Hilbert space, let $B(\mathcal{H})$ be the set of bounded linear operators over $\mathcal{H}$. If $\dim \mathcal{H}<\infty$ we will take $\mathcal{H}=\mathbb{C}^n$, and in that case $B(\mathbb{C}^n)=M_n(\mathbb{C})$. A matrix $A:\mathbb{C}^n\to\mathbb{C}^n$ is {\bf positive}, denoted by $A\geq 0$, if $\langle Av,v\rangle\geq 0$, for all $v\in\mathbb{C}^n$. This is also equivalent to any of the following facts: a) all its eigenvalues are nonnegative; b) there exists a matrix $B$ such that $A=BB^*$. In particular, every positive matrix is hermitian. We say $\rho\in M_n(\mathbb{C})$ is a {\bf density matrix} if $\rho\geq 0$ and $tr(\rho)=1$.

\bigskip

Let $\Phi: M_n(\mathbb{C})\to M_n(\mathbb{C})$ be linear. We say $\Phi$ is a {\bf positive} operator whenever $A\geq 0$ implies $\Phi(A)\geq 0$. In physics problems, we are mostly interested in $\Phi$ that are not only positive on $M_n$ (system A), but also that $I\otimes \Phi$ is also positive, where $I$ is the identity operator on any other choice of component (system B). More precisely, define for each $k\geq 1$, $\Phi_k: M_k(M_n(\mathbb{C}))\to M_k(M_n(\mathbb{C}))$,
\begin{equation}
\Phi_k(A)=[\Phi(A_{ij})],\;\;\; A\in M_k(M_n(\mathbb{C})), \; A_{ij}\in M_n(\mathbb{C})
\end{equation}
We say $\Phi$ is {\bf $k$-positive} if $\Phi_k$ is positive, and we say $\Phi$ is completely positive (CP) if $\Phi_k$ is positive for every $k=1, 2, 3,\dots$. It is well-known that CP maps can be written in the Kraus form \cite{petz}:
\begin{equation}
\Phi(\rho)=\sum_i V_i\rho V_i^*
\end{equation}
Conversely, any such operator is CP. An operator of the form
\begin{equation}
\Phi(\rho)=\sum_{i=1}^k p_i U_i\rho U_i^*,\;\;\; p_i\geq 0, \sum_i p_i=1
\end{equation}
where the $U_i$ are unitary is called a {\bf mixed-unitary operator} or a {\bf random unitary operator}. Such operators are unital ($\Phi(I)=I$) and trace-preserving. It is well-known that $\rho$ is fixed for a mixed-unitary operator if and only if it is fixed for each $F_i(\rho)=U_i\rho U_i^*$ \cite{kribs, novotny}.

\subsection{$\omega$-limit sets}\label{omegalims}

Let $D(\mathbb{C}^N)$ denote the compact, convex space of density matrices over $\mathbb{C}^N$. We identify it with the space of state functionals over $\mathbb{C}^N$. We consider a CPT (trace-preserving) map $\Phi$ acting on $D(\mathbb{C}^N)$. Define, for $\rho\in D(\mathbb{C}^N)$, the {\bf orbit} through $\rho$,
\begin{equation}
\gamma(\rho)=\cup_{n\geq 1} \Phi^n(\rho),
\end{equation}
and the {\bf $\Omega$-limit set} of $\rho$:
\begin{equation}
\Omega(\rho)=\{\eta\in D(\mathbb{C}^N): \eta=\lim_{i\to\infty}\Phi^{n_i}(\rho), n_i\to\infty \textrm{ as } i\to\infty\},
\end{equation}
Define the {\bf attractor} of $\Phi$ to be $\Omega_\Phi=\cup_\rho \Omega(\rho)$. Other definitions are the following.

\bigskip

{\bf Definition.}
\begin{enumerate}
\item We say that a subset $B\subset D(\mathbb{C}^N)$ is {\bf positive invariant} under $\Phi$ if $\Phi^n(B)\subset B$ for all $n\geq 1$.
\item We say that $B$ is {\bf minimal} under $\Phi$ if for any $\eta\in B$, $B=\ov{\gamma(\eta)}$, that is, if $B$ is the closure of the orbit of any of its elements.

\item We say that $B$ is {\bf strongly invariant} under $\Phi$ if for every $n\geq 1$ we have that $\Phi^n$ is a homeomorphism of $B$ onto $B$.
\end{enumerate}

We describe some results which allows us to compare convex sets in the classical setting with the corresponding objects after the quantization procedure we will perform. Most steps are due to results proved in the setting of non-expansive dynamical systems. The main ideas are well-known \cite{dafermos, lemmens}, but for completeness we describe the proof and simple adaptations needed for our setting in the appendix. Our analysis is mostly relevant for CP maps acting on several qubits where, for instance, we obtain periodic asymptotic dynamics, such as the two-qubit CNOT channel \cite{novotny}. When the asymptotic limit is trivial (e.g. amplitude damping channel for one qubit, which has the north pole of the Bloch sphere as limit for all density matrices), many of the results presented are trivial.

\begin{pro}\label{daf_prop1}
For all $\rho\in D(\mathbb{C}^N)$ we have the following. a) $\Omega(\rho)$ is minimal under $\Phi$.
b) For every $k\geq 1$, $\Phi^k$ is an isometry on $\Omega(\rho)$.
c) $\Omega(\rho)$ is strongly invariant under $\Phi$.
\end{pro}

Recall that the {\bf convex hull} of a set $S$, denoted $co\; S$, consists of all the convex combinations of points from that set. Also recall that by Carath\'eodory's theorem, we have that in $d$-dimensional Euclidean space, every point in $co\; S$ is a convex combination of $d+1$ points from $S$. The convex hull of a finite set of points is called a {\bf polytope}. We will be interested in the convex hull for omega-limit sets associated to CP maps acting on finite-dimensional spaces.

\begin{lemma}\label{edel_prop1}
If the restriction of $\Phi$ to a subset $B$ of $D(\mathbb{C}^N)$ is an isometry then the restriction of $\Phi$ to $co\; B$ is also an isometry.
\end{lemma}

\begin{pro}\label{conv_pro1}
a) The set $\ov{co\; \Omega(\rho)}$ is strongly invariant under $\Phi$. Also there is an affine group of isometries on the closed linear variety spanned by $\Omega(\rho)$ which coincides with $\Phi$ on $\ov{co \;\Omega(\rho)}$.
b) $\ov{co\; \Omega(\rho)}$ contains exactly one fixed point $\rho_0$ of $\Phi$, given by
\begin{equation}
\rho_0=\lim_{k\to\infty} \frac{1}{k}\sum_{m=0}^k\Phi^m(\eta), \;\;\;\forall \eta\in \ov{co\;\Omega(\rho)}
\end{equation}
\end{pro}

\begin{pro}\label{lvgaans}
a) There exists a non-expansive projection $\tau$ onto $\Omega_\Phi$. b) We have that $\Omega_\Phi$ is convex and $\Phi$ restricted to such set is an isometry.
\end{pro}

\subsection{On the isometric dynamics}

Whenever we have conjugate dynamics (i.e. a commuting diagram), we are allowed to carry certain properties from the classical to the quantum setting. For instance, we note that for $\Lambda, \kappa_\mu$ and $T$ given in the introduction, all being linear, are such that convex sets are taken to convex sets by these maps, and the same is true for its respective inverse images. Also supposing the commuting diagram it is straightforward to show that if $f$ is a fixed point for $T$ then $\kappa_\mu(f)$ is a fixed point for $\Lambda$. We have seen in propositions \ref{daf_prop1}-\ref{lvgaans} that in certain regions we have an isometric dynamics for $\Lambda$, but such fact is not a consequence of the classical behavior. Instead, this is intrinsic to CP dynamics, which are always non-expansive with respect to the trace distance.

\bigskip

{\bf Example.} Consider the 2-qubit CNOT operator, for which an asymptotic analysis is made in \cite{novotny}. Consider $\Phi(\rho)=C_1\rho C_1 + C_2\rho C_2$, where $C_1|i,j\rangle=|i,i\oplus j\rangle$, $C_2|i,j\rangle=|i\oplus j, j\rangle$. It can be shown that
\begin{equation}
\sigma_1=\lim_{n\to\infty} \rho(2n)=\begin{bmatrix} a & c & c & c \\ \ov{c} & b & d & \ov{d} \\ \ov{c} & \ov{d} & b & d \\ \ov{c} & d & \ov{d} & b\end{bmatrix} ,\;\;\;\; \sigma_2=\lim_{n\to\infty} \rho(2n+1)=\begin{bmatrix} a & c & c & c \\ \ov{c} & b & \ov{d} & d \\ \ov{c} & d & b & \ov{d} \\ \ov{c} & \ov{d} & d & b  \end{bmatrix}
\end{equation}
where $\rho(k):=\Phi^k(\rho)$ and $a, b, c, d$ are numbers depending on the initial state $\rho$ only. A routine calculation shows that $[\sigma_1,\sigma_2]=0$ and that the same holds for any two elements on the segment connecting $\sigma_1$ and $\sigma_2$ (such segment is, of course, $co \{\sigma_1,\sigma_2\}$).

\bigskip

More generally, does commutativity on $\ov{co\; \Omega(\rho)}$ holds for every CP map considered, or there exists a dependence on the classical dynamics and/or the quantization choice? Motivated by the above example, we have that the following holds:

\begin{pro}
a) Suppose that the diagram (\ref{cdd}) commutes for a certain choice of $\mu$ and $f\in L^\infty$. If $g\in L^\infty$ and $f\in\Omega(g)$ then $\kappa_\mu(f)\in\Omega_\Lambda(\kappa_\mu(g))$. b) $[\rho,\eta]=0$ for every $\rho, \eta\in \ov{co \;\Omega(\zeta)}$, $\forall \zeta\in D(\mathbb{C}^N)$.
\end{pro}
{\bf Proof.} a) Immediate from the commutative diagram. b) Let $\rho, \eta\in \ov{co\; \Omega(\zeta)}$. Then
\begin{equation}
\rho=\lim_m \rho_m,\;\;\;\rho_m\in co\; \Omega(\zeta),\;\;\;\rho_m=\sum_{i} \lambda_{m,i} \rho_{m,i},\;\;\; \rho_{m,i}=\lim_j \Phi^{t_{m,i,j}}(\zeta)
\end{equation}
\begin{equation}
\eta=\lim_n \eta_n,\;\;\;\eta_n\in co\; \Omega(\zeta),\;\;\;\eta_n=\sum_{k} \lambda_{n,k} \eta_{n,k},\;\;\; \eta_{n,k}=\lim_l \Phi^{t_{n,k,l}}(\zeta)
\end{equation}
Therefore,
\begin{equation}
\rho=\lim_m \sum_{i} \lambda_{m,i} \lim_j \Phi^{t_{m,i,j}}(\zeta),\;\;\;\; \eta=\lim_n \sum_{k} \lambda_{n,k}\lim_l \Phi^{t_{n,k,l}}(\zeta)
\end{equation}
The conclusion follows.

\qed

{\bf Remark. } It is not true that every two densities in the attractor set of a CP $\Phi$ commute, as simple examples can show. However, if they belong to the convex hull of the same density matrix, then this is true as the above proposition shows.

\section{Quantization procedure}\label{qua_proc}

\subsection{Measures and a representation for $M_n$}\label{meas_ortho}

Measure theory has many applications to algebras appearing in quantum mechanics. Some measure-theoretic analysis that occur in quantum information problems are related to identifying properties of states. For instance, entanglement of formation can be described in terms of integrals over the state space \cite{majewski}. In order to introduce our discussion, we recall the following definition. A {\bf barycenter} of a probability measure $\mu\in M_1(K)$ over a compact, convex set $K$ is an element $b(\mu)\in K$ such that $\psi(b(\mu))=\int \psi(\omega)\; d\mu(\omega)$,  for every $\psi$ linear functional on $K$. Two basic results on barycenters of measures are the following \cite{bratteli}:
\begin{theorem}
Let $K$ be compact and convex, $\mu\in M_1(K)$, the set of probability measures over $K$.

a) There exists a unique point $b(\mu)\in K$ such that
\begin{equation}
\psi(b(\mu))=\int\psi\;d\mu,\;\; \psi\in A(K),
\end{equation}
where $A(K)$ denotes the set of real continuous affine functions.

b) There exists a net $\mu_\alpha\in M_1(K)$ of measures with finite support which is weakly$^*$ convergent to $\mu$ and is such that $b(\mu_\alpha)=b(\mu)$, for all $\alpha$.
\end{theorem}

\begin{theorem}
Let $\mathcal{U}$ be a C$^*$-algebra and $E_\mathcal{U}$ the associated set of states. Then the set $M_\omega(E_\mathcal{U})$ of probability measures over $E_\mathcal{U}$ with barycenter $\omega$ is a convex, compact subset of $M_1(E_\mathcal{U})$, in the weak$^*$ topology.
\end{theorem}

We consider the $C^*$-algebra $M=M_n$ of order $n$ complex matrices. Recall that a state is $\omega\in M_n^*$, $\omega: M_n\to\mathbb{C}$ such that it is positive (i.e., takes positive elements to positive real numbers, $\omega(AA^*)\geq 0$) and $\omega(I)=1$. Let $\rho\in M_n$ be a faithful state and identify its GNS triplet $(H_\rho,\pi_\rho,\Omega_\rho)$ with
\begin{equation}
(\mathbb{C}^n\otimes\mathbb{C}^n, M_n\otimes I_n,|\sqrt{\rho}\raa)
\end{equation}
That is, $\pi_\rho(M_n)=M_n\otimes I_n$, where if $\rho=\sum_i r_i |r_i\raa\laa r_i|\in M_n$ then
\begin{equation}
|\sqrt{\rho}\raa=\sum_i \sqrt{r_i}|r_i\raa\otimes |r_i\raa\in \mathbb{C}^n\otimes\mathbb{C}^n
\end{equation}
Also the associated state functional is $\omega_\rho:M_n\to\mathbb{C}$, $\omega_\rho(X)=tr(X\rho)$. It also holds that the linear span of vectors of the form $|X\sqrt{\rho}\raa$, $X\in M_n$, is dense in $\mathbb{C}\otimes\mathbb{C}$. And note that $\pi_\rho(M_n)'=I_n\otimes M_n$.

\bigskip

{\bf Remark.} In principle we could have chosen a representation that takes $L^\infty$ to $M_n$, instead of $M_n\otimes I_n$. We prefer to use the latter because this representation makes more explicit the relation between the finite algebra we are considering and its commutant. In a general setting, for a given $C^*$-algebra $\mathcal{A}$, $\kappa_\mu$ takes $L^\infty$ to the commutant of $\pi(\mathcal{A})$. This particular representation also appears in the modular theory for finite-level systems \cite{benatti}.

\bigskip

{\bf Example}. (Orthogonal measures \cite{bratteli,majewski, takesaki}) Denote by $E_M$ the state space for $M=M_n$, the C$^*$-algebra of matrices. Recall that $E_M$ is a convex and compact in the weak$^*$ topology. Denote by $\mathcal{E}(E_M)$ the set of extreme points of $E_M$. For $\rho\in E_M$, denote by $M_\rho(E_M)$ the space of probability measures with barycenter $\rho$. Let $\omega_1$ and $\omega_2$ be positive linear functionals over $M_n$ and let $\omega=\omega_1+\omega_2$. We say that $\omega_1$ is {\bf orthogonal} to $\omega_2$, written $\omega_1\perp \omega_2$ if there exists a projection $P\in I_n\otimes M_n$ such that
\begin{equation}
\omega_1(B)=\langle P\Omega_\omega,(B\otimes I)\Omega_\omega\rangle,\;\;\;\omega_2(B)=\langle (I-P)\Omega_\omega,(B\otimes I)\Omega_\omega\rangle
\end{equation}
Let $\mu$ be a positive regular Borel measure on $E_M$. We say $\mu$ is an {\bf orthogonal} measure if
\begin{equation}
\int_S \omega\; d\mu(\omega) \perp \int_{E_M\setminus S} \omega\; d\mu(\omega)
\end{equation}
for every $S\subset E_M$ Borel set. We note that orthogonal measures are in one-to-one correspondence with the abelian von Neumann subalgebras of the commutant $\pi_\rho(M)'$ \cite{bratteli}.

\subsection{Transfer operator}\label{tomita}

With the notation of the previous section, let $E_M$ denote the convex weakly$^*$ compact set of states over $M$. Let $M_\omega(E_M)$ be the set of probability measures over $E_M$ with barycenter $\omega$. For $\mu\in M_\omega(E_M)$ define the map $f \in L^\infty(E_M;\mu)\mapsto \kappa_\mu(f)\in \pi_\rho(M)'$,
\begin{equation}
\langle\Omega_\omega,\kappa_\mu(f)\pi_\rho(A)\Omega_\omega\rangle=\int f(\omega')\omega'(A)\; d\mu(\omega'),
\end{equation}
which is positive and non-expansive. We know by Tomita's theorem \cite{bratteli} that if $\mu$ is a nonnegative regular orthogonal Borel measure on $E_M$ then the map $f \mapsto \kappa_\mu(f)$ is a $*$-isomorphism of $L^\infty(E_M;\mu)$ into $\pi_\rho(M)'$ and
\begin{equation}
\mathcal{A}=\{\kappa_\mu(f): f \in L^\infty(E_M;\mu)\}
\end{equation}
is an abelian von Neumann subalgebra of $\pi_\rho(M)'$. Now define $L: L^\infty(E_M;\mu)\to L^\infty(E_M;\mu)$,
\begin{equation}
L(f)(A)=\sum_i p_if(F_i(A)),
\end{equation}
where $F_i: M_n\to M_n$, $A\in M_n$, $f \in E_M$, $p_i\geq 0$. Such operator is known as Ruelle {\bf transfer operator}, or Ruelle-Perron-Frobenius operator. The spectral theory of this map is known as thermodynamic formalism \cite{baladi}.

\bigskip

{\bf Remark.} When studying transfer operators, it is usual to consider them acting on the space $C(E_M)$ of continuous functions. The analysis made in this article also works in that setting, except for the statements related to orthogonal measures and algebra commutants. Our preference for the more general space $L^\infty$ is due to the fact that in such space one benefits from certain algebraic properties encountered when studying the operator $\kappa_\mu$ in full generality, see \cite{bratteli}.

\bigskip

While the interplay between the transfer operator $L$ (acting on continuous functions) and its dual $L^*$ (acting on measures) has been extensively studied \cite{baladi}, little is known about the behavior of the noncommutative counterparts of the measures (the state operators) with respect to quantum analogues of the transfer operator (acting on matrix algebras). We aim to describe a relation between $L$ acting on $L^\infty$ functions and completely positive maps (quantum channels) acting on a matrix algebra. Note that $L$ itself is completely positive, since it is a positive operator acting on a commutative $C^*$-algebra.

\subsection{Conditions on the CP map}\label{conditions_on}

We are interested in the evolution of $L(f)$ and its relation to $\kappa_\mu(f)$:
\begin{equation}
\begin{CD}L^\infty(E_M;\mu) @>L>> L^\infty(E_M;\mu) @>L>> \cdots \\
@V{\kappa_\mu}VV  @V{\kappa_\mu}VV \\
I_n\otimes M_n @>{\Lambda}>> I_n\otimes M_n @>{\Lambda}>> \cdots
\end{CD}
\end{equation}
where $\Lambda$ is a CP map to be determined.

\bigskip

{\bf Definition.} We say that $\kappa_\mu$ is a {\bf CP quantization} for the transfer map $T: L^\infty\to L^\infty$ if there exists $\Lambda$ CP such that the above diagram commutes. We say the quantization is {\bf faithful} if $f\in L^\infty(\mu)\mapsto \kappa_\mu(f)\in I_n\otimes M_n$ is faithful as a representation. It is well-known that a quantization is faithful if and only if $\mu\in\mathcal{E}(M_\omega(E_M))$ \cite{bratteli}.

\bigskip

We are interested in finding conditions over $\Lambda$ in order to obtain a commutative diagram, that is, $\kappa_\mu\circ L=\Lambda\circ \kappa_\mu$. We will suppose that $\Lambda:I_n\otimes M_n\to I_n\otimes M_n$ is completely positive. Note that $I\otimes M_n\simeq I_n\otimes B(\mathbb{C}^n)\subset B(\mathbb{C}^n\otimes\mathbb{C}^n)\simeq B(\mathbb{C}^{n^2})$. We will then take $V_i: \mathbb{C}^{n}\to \mathbb{C}^{n}$ such that
\begin{equation}
\Lambda(I_n\otimes X)=\sum_i (I_n\otimes V_i)(I_n\otimes X)(I_n\otimes V_i^*)
\end{equation}
We have
\begin{equation}
\langle \Omega_\rho,\kappa_\mu(Lf)\pi_\rho(A)\Omega_\rho\rangle=\langle \Omega_\rho,(I\otimes M_{Lf })\pi_\rho(A)\Omega_\rho\rangle=\int \sum_i p_if(F_i(\omega'))\omega'(A)\; d\mu(\omega'),
\end{equation}
where above we write $\kappa_\mu(Lf)=I\otimes M_{Lf }$. On the other hand,
$$\langle \Omega_\rho,\Lambda(\kappa_\mu(f))\pi_\rho(A)\Omega_\rho\rangle=\langle \Omega_\rho,\Lambda(I\otimes M_f)\pi_\rho(A)\Omega_\rho\rangle$$
\begin{equation}
=\langle\Omega_\rho,\sum_i (I_n\otimes V_i)(I_n\otimes M_f)(I_n\otimes V_i^*)\pi_\rho(A)\Omega_\rho\rangle=\langle\Omega_\rho,\sum_i I_n\otimes (V_iM_f  V_i^*)\pi_\rho(A)\Omega_\rho\rangle
\end{equation}
Therefore we are interested in conditions over the $V_i$ in order to obtain
\begin{equation}
I_n\otimes M_{Lf }=I_n\otimes \sum_i V_iM_f  V_i^*
\end{equation}
In order to find a solution to the above equation, it is enough to solve
\begin{equation}\label{eq_a_resolver}
M_{Lf }=\sum_i V_iM_f  V_i^*
\end{equation}

Now we perform the general procedure and for clarity the reader may consider at each step the case $n=2$. Let $\mu\in M_{\rho}(E_{M_n})$, the set of measures over the state space for $M_n$, with barycenter $\rho$. Let $f \in L^\infty(\mu)$. Then $\kappa_\mu(f)\in \pi_\rho(M_n(\mathbb{C}))'=I_n\otimes M_n(\mathbb{C})$. It is a simple calculation to show that $\{\pi_\rho(E_{ij})|\sqrt{\rho}\raa\}_{i,j=1,.\dots n}$ is a basis for $\mathbb{C}^n\otimes\mathbb{C}^n$, where $E_{ij}$ are the matrix units for $M_n$. For instance, if $n=2$, write $|0\raa=(1,0)$, $|1\raa=(0,1)$, $|\sqrt{\rho}\raa=p_0|0\raa\otimes|0\raa+p_1|1\raa\otimes|1\raa$. Then
$$\pi_\rho(E_{11})|\sqrt{\rho}\raa=(E_{11}\otimes I_2)(p_0|0\raa\otimes|0\raa+p_1|1\raa\otimes|1\raa)$$
\begin{equation}
=p_0(E_{11}\otimes I_2)(|0\raa\otimes|0\raa)+p_1(E_{11}\otimes I_2)(|1\raa\otimes|1\raa)=(p_0,0,0,0)^T
\end{equation}
Similarly,
\begin{equation}
\pi_\rho(E_{12})|\sqrt{\rho}\raa=(0,p_1,0,0)^T,\;\;\;\pi_\rho(E_{21})|\sqrt{\rho}\raa=(0,0,p_0,0)^T,\;\;\;\pi_\rho(E_{22})|\sqrt{\rho}\raa=(0,0,0,p_1)^T
\end{equation}

To avoid too many indices we make a pair $(i,j)$ correspond to the number $n(i-1)+j$. That is $11\to 1$, $12\to 2$, $21\to 3$, $22\to 4$ and so on. In this way we relabel the matrix units. Denote $v_{k}=\pi_\rho(E_{k})|\sqrt{\rho}\raa$, for $k=1,\dots,n^2$. In such basis we have
\begin{equation}
[\kappa_\mu(f)]_{ij}=\laa v_i|\kappa_\mu(f)v_j\raa=\laa\sqrt{\rho}|\pi_\rho(E_i)|\kappa_\mu(f)\pi_\rho(E_j)|\sqrt{\rho}\raa
\end{equation}
\begin{equation}\label{partial_c}
=\laa\sqrt{\rho}|\kappa_\mu(f)\pi_\rho(E_i^*)\pi_\rho(E_j)|\sqrt{\rho}\raa=\laa\sqrt{\rho}|\kappa_\mu(f)\pi_\rho(E_i^*E_j)|\sqrt{\rho}\raa=\int f(\omega')\omega'(E_i^*E_j)d\mu(\omega')
\end{equation}
A routine calculation  proves that the matrix $\kappa_\mu(f)$ is block diagonal and all blocks are equal. For instance, if $n=2$ such matrix has order 4 and
$$E_1^*E_1=E_1,\;\; E_1^*E_2=E_2,\;\; E_1^*E_3=0,\;\; E_1^*E_4=0$$
$$E_2^*E_1=E_3,\;\; E_2^*E_2=E_4,\;\; E_2^*E_3=0,\;\; E_2^*E_4=0$$
$$E_3^*E_1=0,\;\; E_3^*E_2=0,\;\; E_3^*E_3=E_1,\;\; E_3^*E_4=E_2$$
$$E_4^*E_1=0,\;\; E_4^*E_2=0,\;\; E_4^*E_3=E_3,\;\; E_4^*E_4=E_4$$
Hence we obtain the matrix
$$\kappa_\mu(f)=\begin{bmatrix} \int f \;\omega(E_1)d\mu & \int f \;\omega(E_2)d\mu & 0 & 0 \\ \int f \;\omega(E_3)d\mu & \int f \;\omega(E_4)d\mu & 0 & 0 \\ 0 & 0 & \int f \;\omega(E_1)d\mu & \int f \;\omega(E_2)d\mu \\ 0 & 0 & \int f \;\omega(E_3)d\mu & \int f \;\omega(E_4)d\mu  \end{bmatrix} $$
\begin{equation}
=I\otimes \begin{bmatrix} \int f \;\omega(E_1)d\mu & \int f \;\omega(E_2)d\mu \\ \int f \;\omega(E_3)d\mu & \int f \;\omega(E_4)d\mu \end{bmatrix}=I\otimes M_f ,
\end{equation}
thus providing an explicit expression for $M_f $, and analogously for $M_{Lf }$. So equation (\ref{eq_a_resolver}) becomes, for $n=2$,
$$M_{Lf }=\sum_i V_i M_f  V_i^* \; \Leftrightarrow$$
\begin{equation}\label{eq_a_resolver_expl}
\sum_{i=1}^k p_i \begin{bmatrix} \int f(F_i)\;\omega(E_1)d\mu & \int f(F_i)\;\omega(E_2)d\mu \\ \int f(F_i)\;\omega(E_3)d\mu & \int f(F_i)\;\omega(E_4)d\mu \end{bmatrix}=\sum_{i=1}^l V_i \begin{bmatrix} \int f \;\omega(E_1)d\mu & \int f \;\omega(E_2)d\mu \\ \int f \;\omega(E_3)d\mu & \int f \;\omega(E_4)d\mu \end{bmatrix}V_i^*
\end{equation}
The number $k$ is determined by the transfer operator and can be at most $n^2$, a restriction obtained from the Kraus decomposition. To solve the above equation, it is enough to look for $V_i$ which are diagonal, and we also set $l=k$. For a general matrix $A=(a_{ij})$, we have
\begin{equation}
V_i=\textrm{diag}(v_1^i,\dots,v_n^i) \Rightarrow\; V_iAV_i^*=[v_r^i\ov{v_s^i}a_{rs}],\;\; r,s=1,\dots, n, \;\;i=1,\dots, k
\end{equation}
Now assume a correspondence between the $i$-th matrix on each side of equation (\ref{eq_a_resolver_expl}). Then for the diagonal $V_i$ above and
\begin{equation}
A=M_f =[\int f \;\omega(E_{rs})d\mu],\;\;\; r,s=1,\dots, n
\end{equation}
we get from (\ref{eq_a_resolver_expl}),
\begin{equation}
p_i [\int f(F_i)\;\omega(E_{rs})d\mu]= v_r^i\ov{v_s^i}\int f \;\omega(E_{rs})d\mu,\;\;\; r,s=1,\dots, n,\;\;\; i=1,\dots, k
\end{equation}
For $n=2$, this becomes
\begin{equation}
p_i \begin{bmatrix} \int f(F_i)\;\omega(E_1)d\mu & \int f(F_i)\;\omega(E_2)d\mu \\ \int f(F_i)\;\omega(E_3)d\mu & \int f(F_i)\;\omega(E_4)d\mu \end{bmatrix}=\begin{bmatrix} |v_1^i|^2 \int f \;\omega(E_1)d\mu & v_1^i\ov{v_2^i}\int f \;\omega(E_2)d\mu \\ \ov{v_1^i}v_2^i \int f \;\omega(E_3)d\mu & |v_2^i|^2\int f \;\omega(E_4)d\mu\end{bmatrix}
\end{equation}
Note that the equations obtained from the diagonal are
\begin{equation}
p_i\int f(F_i(\omega))\omega(E_{rr})d\mu(\omega)=|v_{r}^i|^2\int f(\omega)\omega(E_{rr})d\mu(\omega),\;\;\;i=1,\dots, k
\end{equation}
and so
\begin{equation}
|v_{r}^i|=\sqrt{\frac{p_i\int f(F_i(\omega))\omega(E_{rr})d\mu(\omega)}{\int f(\omega)\omega(E_{rr})d\mu(\omega)}},\;\;\; r=1,\dots, n,\;\;\;i=1,\dots, k
\end{equation}
On the other hand the equations obtained from non-diagonal entries must also satisfy such conditions, but this is not expected to hold in general. So we have restrictions on the states with respect to a given measure $\mu$.

\bigskip

{\bf Example.} Take $n=2$. If we choose $\mu=\delta_{\omega_0}$, we obtain the equations
\begin{equation}
p_if(F_i(\omega_0))=|v_1^i|^2f(\omega_0)
\end{equation}
\begin{equation}
p_if(F_i(\omega_0))=v_1^i\ov{v_2^i}f(\omega_0)
\end{equation}
\begin{equation}
p_if(F_i(\omega_0))=|v_2^i|^2f(\omega_0)
\end{equation}
So if $f(\omega_0)\neq 0$ we conclude
\begin{equation}
|v_1^i|^2=|v_2^i|^2=p_i\frac{f(F_i(\omega_0))}{f(\omega_0)}, \; \; i=1,\dots, k
\end{equation}
Note that if $\omega_0$ is a fixed point for the CP map $\Lambda: \rho\mapsto \sum_i p_iF_i$ then
\begin{equation}
v_1^i=v_2^i=\sqrt{p_i}e^{i\theta_i}, \; \; i=1,2, \;\; \theta_i\in\mathbb{R}
\end{equation}
This establishes the existence of a solution for the commutative diagram described above in the present case. Note that if $\sum_i p_i=1$, as it is usually assumed, then $\Lambda(I)=V_1V_1^*+V_2V_2^*=I$ so we have that if $\omega_0$ is a fixed point for $\Lambda$ which is mixed-unitary then each $F_i$ is fixed by such point and therefore $\Lambda$ is unital and trace-preserving (see section \ref{cp}).

\bigskip

An open question is to find all solutions for a general quantum channel, i.e., in the case that not all $V_i$ are diagonal. Also the case in which $\mu$ is a convex combination of Dirac deltas is of interest, a problem where a generalization of the above method seems possible. Another open question is to describe the solution when we consider an orthogonal measure.

\bigskip

{\bf Example.} Consider $M_2(\mathbb{C})$, let $\mu=\lambda_1\delta_{\omega_1}+\lambda_2\delta_{\omega_2}$, for $\omega_1$, $\omega_2$ fixed states with support on the diagonal units of $M_2$. Write $\kappa_\mu(Lf)=I\otimes M_{Lf}$, $\kappa_\mu(f)=I\otimes M_{f}$. Then we wish to obtain $V_j$ satisfying $M_{Lf}=\sum_j V_jM_f V_j^*$, and again we seek $V_j$ which are diagonal. For each $i$, we have:
\begin{equation}
p_i \int f(F_i(\omega))\;\omega(E_{rs})d\mu= v_r^i\overline{v_s^i}\int f(\omega) \;\omega(E_{rs})d\mu,\;\;\; r,s=1,\dots, n,\;\;\; i=1,\dots, k
\end{equation}
We conclude that
\begin{equation}
|v_1^i|^2=\frac{p_i\Big(\lambda_1 f(F_i(\omega_1))\omega_1(E_{11})+\lambda_2 f(F_i(\omega_2))\omega_2(E_{11})\Big)}{\lambda_1 f(\omega_1)\omega_1(E_{11})+\lambda_2 f(\omega_2)\omega_2(E_{11})}
\end{equation}
\begin{equation}
|v_2^i|^2=\frac{p_i\Big(\lambda_1 f(F_i(\omega_1))\omega_1(E_{22})+\lambda_2 f(F_i(\omega_2))\omega_2(E_{22})\Big)}{\lambda_1 f(\omega_1)\omega_1(E_{22})+\lambda_2 f(\omega_2)\omega_2(E_{22})}
\end{equation}
The above reasoning can be generalized and summarized in the following:

\begin{theorem}\label{the_theorem}
Let $L: C(E_{M_n})\to C(E_{M_n})$, $L(g)(\rho)=\sum_{i=1}^k p_i g(F_i(\rho))$, $k\leq n^2$, $p_i\geq 0$.
\begin{enumerate}

\item For a given $f>0$ and for every $\mu=\sum_m\lambda_m\delta_{\omega_m}$, $\lambda_m\geq 0$, $\omega_i$ with support on the diagonal units of $M_n(\mathbb{C})$, there exists a solution for $\Lambda\circ \kappa_\mu(f)=\kappa_\mu(Lf)$, given by $\Lambda(\rho)=\sum_{i=1}^k V_i\rho V_i^*$, $V_i=\textrm{diag}(v_1^i,\dots, v_n^i)$, and
\begin{equation}
|v_j^i|=\sqrt{\frac{p_i\Big(\sum_m\lambda_m f(F_i(\omega_m))\omega_m(E_{jj})\Big)}{\sum_m\lambda_m f(\omega_m)\omega_m(E_{jj})}},\;\;\; i=1,\dots, k,\;\;\; j=1,\dots,n
\end{equation}

\item If $Lf=\sum_i p_i f(F_i)$ with $F_i=U_i\cdot U_i^*$, $U_i$ unitary, there exists $V_i$ which satisfies $\Lambda\circ \kappa_\mu(f)=\kappa_\mu(Lf)$ for every $f\in L^\infty$ if and only if the $\omega_i$ are all fixed points for $\Phi(\rho)=\sum_i p_iF_i(\rho)$.
\end{enumerate}
\end{theorem}

\section{Classical diagram and open questions}\label{conclusion}

We recall that a density matrix $\rho_0$ is a fixed point for a mixed unitary channel $\Lambda$ if and only if such matrix is the barycenter of a measure which is invariant for the associated Markov operator $P$. If $\Lambda(\rho)=\sum_i p_i F_i(\rho)$, $F_i(\rho)=U_i\rho U_i^*$ then we know that $\Lambda(\rho_0)=\rho_0$ implies that $F_i(\rho_0)=\rho_0$ for each $i$ \cite{novotny}. The associated Markov operator $P_M$ on measures is given by
\begin{equation}\label{markovop}
P_M\mu(B)=\sum_i \int_{F_i^{-1}(B)} p_i\; d\mu
\end{equation}
Classical Markov operators are related to the discussion in this paper in the following way. Consider the following diagram:
\begin{equation}
\begin{CD}
M_a @>P_{M}>> M_a \\
@A{I}AA  @V{R}VV \\
L^\infty(E_M;\mu) @>L, P>> L^\infty(E_M;\mu) \\
@V{\kappa_\mu}VV  @V{\kappa_\mu}VV \\
I_n\otimes M_n @>{\Lambda}>> I_n\otimes M_n
\end{CD}
\end{equation}
The lower part of the diagram is the one considered in the previous sections. Above, $M_a$ denotes the family of absolutely continuous (AC) measures, $I$ is the integral formula map $f\mapsto \mu_f$, $\mu_f(A)=\int_A f\;d\mu$ and $R$ stands for the Radon-Nikodym derivative of an AC measure. Assume that $P_M$ is a Markov operator on measures acting on absolutely continuous measures. Now let $f\in L^1$, $f\geq 0$ and consider $\mu_f$. Then $P_M(\mu_f)$ is also AC, so we can write it in the form $P_M\mu_f(A)=\int_A g(\rho)\;d\mu(\rho)$, where $g$ is the Radon-Nikodym derivative with respect to the Borel measure $P_M\mu_f$ on $D(\mathbb{C}^n)$. Therefore for every $f\in L^1$ we obtain a unique $g\in L^1$, such correspondence is a Markov operator $P$ acting on densities (positive functions) and it is such that the upper part of the above diagram (not involving matrices) commutes \cite{lasota}.

\bigskip

On the other hand, suppose we are given $\Lambda$ which can be written as $\Lambda(\rho)=\sum_i p_i F_i(\rho)$. Then we can obtain a Markov operator $P_M$ on measures, via equation (\ref{markovop}), and therefore another operator $P$ acting on densities. It is an open question to determine in which cases we have that the $P$ obtained is such that the lower diagram commutes. This is the inverse problem: given a CP map on a matrix algebra, obtain a classical approximation.

\bigskip

Another open question is the following: given $L$ operator on positive functions and $f$ a function, consider a quantization $\Lambda_f$ via $\kappa_\mu$, a Markov operator $P_M$ on measures and by the above construction an operator $P$ on positive functions. Then, we wish to determine how similar are $L$ and $P$, and whether we can find cases in which $L=P$ (or at least $Lf=Pf$), and whether such solutions are nontrivial.

\section{Appendix}

In this appendix, we present the results on non-expansive maps cited in this work. The main ideas have been presented in \cite{dafermos, edelstein, lemmens}, and are adapted here for our purposes. For completeness we present the proofs, which are mostly slight variations of the works mentioned.

\bigskip

{\bf Proposition \ref{daf_prop1}}. \cite{dafermos} For all $\rho\in D(\mathbb{C}^N)$ we have the following. a) $\Omega(\rho)$ is minimal under $\Phi$. b) For every $k\geq 1$, $\Phi^k$ is an isometry on $\Omega(\rho)$. c) $\Omega(\rho)$ is strongly invariant under $\Phi$.

\bigskip

{\bf Proof.} a) Let $\eta\in\Omega(\rho)$, $\eta=\lim_{k\to\infty} \Phi^{n_k}(\rho)$, $n_k\to\infty$ as $k\to\infty$. For any $l\geq 1$, $\Phi^l(\eta)=\lim_{k\to\infty}\Phi^{n_k+l}(\rho)\in\Omega(\rho)$. Hence $\ov{\gamma(\eta)} \subseteq\ov{\Omega(\rho)}=\Omega(\rho)$. Now let $\zeta\in\Omega(\rho)$, $\zeta=\lim_{k\to\infty} \Phi^{\tau_k}(\rho)$, $\tau_k\to\infty$ as $k\to\infty$. Without loss of generality assume $s_k:= \tau_k-n_k\geq k$, $k=1, 2,\dots$. Now note that
\begin{equation}
d(\Phi^{s_k}(\eta),\zeta)\leq d(\Phi^{s_k}(\eta),\Phi^{s_k+n_k}(\rho))+d(\Phi^{\tau_k}(\rho),\zeta)\leq d(\eta, \Phi^{n_k}(\rho))+d(\Phi^{\tau_k}(\rho),\zeta)
\end{equation}
so we deduce that $\zeta\in \Omega(\eta)$. Hence we have $\Omega(\rho)\subseteq\Omega(\eta)\subseteq\ov{\gamma(\eta)}$ and so we get $\Omega(\rho)=\Omega(\eta)=\ov{\gamma(\eta)}$. Therefore $\Omega(\rho)$ is minimal.

\bigskip

b) Note that from a), we have that $\Omega(\rho)=\Omega(\eta)$ so there is $\{\sigma_k\}\subset \mathbb{N}$, $\sigma_k\to\infty$ such that $\eta=\lim_{k\to\infty} \Phi^{\sigma_k}(\eta)$. We claim that $\Phi^{\sigma_k}(\zeta)\to\zeta$, $k\to\infty$, for any $\zeta \in\Omega(\rho)$. Indeed, if $\zeta=\lim_{k\to\infty} \Phi^{s_k}(\eta)$ then
$$d(\Phi^{\sigma_k}(\zeta),\zeta)\leq d(\Phi^{\sigma_k}(\zeta),\Phi^{\sigma_k+s_k}(\eta))+d(\Phi^{\sigma_k+s_k}(\eta),\Phi^{s_k}(\eta))+d(\Phi^{s_k}(\eta),\zeta)\leq$$
\begin{equation}
\leq 2 d(\zeta,\Phi^{s_k}(\eta))+d(\Phi^{\sigma_k}(\eta),\eta)\to 0
\end{equation}
as $k\to\infty$. In the last inequality we have applied the non-expansiveness of $\Phi$ and also due to this fact we have for any $k\geq 1$,
\begin{equation}
d(\eta,\zeta)=\lim_{k\to\infty} d(\Phi^{\sigma_k}(\eta),\Phi^{\sigma_k}(\zeta)) \leq d(\Phi^k(\eta),\Phi^k(\zeta)) \leq d(\eta,\zeta)
\end{equation}
This shows that $\Phi^k$ is an isometry on $\Omega(\rho)$.

\bigskip

c) From b) we get that  $\Phi^k: \Omega(\rho)\to\Omega(\rho)$ is one-to-one. To prove that $\Phi^k$
is onto $\Omega(\rho)$, we fix $\zeta\in\Omega(\rho)$. For large $l, m, n,$
$$d(\Phi^{\sigma_n-k}(\zeta),\Phi^{\sigma_m-k}(\zeta)) \leq$$
$$\leq d(\Phi^{\sigma_n-k}(\zeta),\Phi^{\sigma_n+\sigma_l-k}(\zeta))+d(\Phi^{\sigma_n+\sigma_l-k}(\zeta),\Phi^{\sigma_m+\sigma_l-k}(\zeta))+d( \Phi^{\sigma_m+\sigma_l-k}(\zeta),\Phi^{\sigma_m-k}(\zeta))\leq $$
\begin{equation}
\leq 2d(\Phi^{\sigma_l}(\zeta),\zeta)+d(\Phi^{\sigma_n+\sigma_l-k}(\zeta),\Phi^{\sigma_m+\sigma_l-k}(\zeta))
\end{equation}
Also from b) we have that $\Phi^{\sigma_l}(\zeta)\to\zeta$, $l\to\infty$. Moreover, for fixed l, $\{\Phi^{\sigma_n+\sigma_l-k}(\zeta)\}$
is Cauchy, converging to $\Phi^{\sigma_l-k}(\zeta)$ as $n\to\infty$. It follows that
$\{\Phi^{\sigma_n - k}(\zeta)\}$ is Cauchy, so let $\eta$ denote its limit. Then it is clear that
$\Phi^k(\eta)= \lim_{n\to\infty}\Phi^{\sigma_n}(\zeta)= \zeta$, so  $\Phi^k$ is onto. This proves the strong invariance of $\Omega(\rho)$.

\qed

{\bf Lemma \ref{edel_prop1}.} \cite{edelstein} If the restriction of $\Phi$ to a subset $B$ of $D(\mathbb{C}^N)$ is an isometry then the restriction of $\Phi$ to $co\; B$ is also an isometry.

\bigskip

{\bf Proof.} To prove that $d(\Phi(\rho),\Phi(\eta))=d(\rho,\eta)$, for all $\rho, \eta\in co\; B$,  it is enough to prove the equality for $\rho_1\in co\; B$ and $\rho\in B$:
\begin{equation}\label{part_conv1}
d(\Phi(\rho),\Phi(\eta))=d(\rho,\eta),\;\;\; \rho_1\in co\; B, \eta\in B
\end{equation}
In fact if this is true then $\Phi$ restricted to $B_1=B\cup\{\rho_1\}$ is an isometry so the above equality holds for $\rho_1\in B_1$ (but also $\rho_1\in co\; B$) and $\rho_2\in co\; B_1=co(B\cup \{\rho_1\})=co\; B$. To prove (\ref{part_conv1}) we use induction on the positive integer $k$ in the representation $\rho=\sum_{i=1}^k \lambda_i\rho_i$, $\lambda_i>0$, $\sum_{i=1}^k\lambda_i=1$, $\rho_i\in B$. For $k=1$, the claim is true by assumption. Suppose it is true for $k-1$ and write
\begin{equation}
\rho=\sum_{i=1}^{k-1}\lambda_i\rho_i+\lambda_k\rho_k=\alpha \rho_1+(1-\alpha)\rho_2
\end{equation}
where
\begin{equation}
\alpha=\sum_{i=1}^{k-1}\lambda_i,\;\;\;\rho_1=\frac{1}{\alpha}\sum_{i=1}^{k-1}\lambda_i\rho_i,\;\;\; \rho_2= \rho_k
\end{equation}
Now note that $\rho_2\in B$ and $\rho_1$ satisfies (\ref{part_conv1}) by induction hypothesis. Therefore the triangles $\eta\rho_1\rho_2$ and $\Phi(\eta)\Phi(\rho_1)\Phi(\rho_2)$ are congruent. In particular $d(\Phi(\rho_1),\Phi(\rho_2))=d(\rho_1,\rho_2)$. Since $\rho$ belongs to the segment generated by $\rho_1$ and $\rho_2$ we have
\begin{equation}
d(\rho,\rho_1)=d(\rho_1,\rho_2) - d(\rho,\rho_2)\leq d(\Phi(\rho_1),\Phi(\rho_2)) - d(\Phi(\rho),\Phi(\rho_2)) \leq d(\Phi(\rho_1),\Phi(\rho))\leq d(\rho_1,\rho)
\end{equation}
It follows that $\eta\rho_1\rho$ and $\Phi(\eta)\Phi(\rho_1)\Phi(\rho)$ are congruent. So (\ref{part_conv1}) holds and this concludes the lemma.

\qed

{\bf Proposition \ref{conv_pro1}.}\cite{dafermos} a) The set $\ov{co\; \Omega(\rho)}$ is strongly invariant under $\Phi$. Also there is an affine group of isometries on the closed linear variety spanned by $\Omega(\rho)$ which coincides with $\Phi$ on $\ov{co \;\Omega(\rho)}$.
b) $\ov{co\; \Omega(\rho)}$ contains exactly one fixed point $\rho_0$ of $\Phi$, given by
\begin{equation}
\rho_0=\lim_{k\to\infty} \frac{1}{k}\sum_{m=0}^k\Phi^m(\eta), \;\;\;\forall \eta\in \ov{co\;\Omega(\rho)}
\end{equation}
{\bf Proof.} a) The first claim follows from a convexity reasoning over $\Omega(\rho)$, which is strongly invariant: the surjective argument is immediate and injectivity follows from the fact that $\Phi^k$ is an isometry on $co\; \Omega(\rho)$ also, by lemma \ref{edel_prop1}. Let $L$ denote the affine space spanned by $\Omega(\rho)$. For $\rho_0\in L$, write $\rho_0=\lambda \eta+(1-\lambda)\zeta$, $\eta, \zeta\in co \;\Omega(\rho)$. Define $T^k(\rho_0):=\lambda\Phi^k(\eta)+(1-\lambda)\Phi^k(\zeta)$, $k\geq 0$. It is straightforward to check that $T^k(\rho_0)$ does not depend on the particular convex expression for $\rho_0$. Also it is clearly affine, satisfying $T^0=I$ and $T^{p+q}=T^pT^q$, $p,q\geq 1$. By continuity we extend the definition to $\ov{L}$.

\bigskip

b) Suppose $\eta_0\in\ov{co\; \Omega(\rho)}$ is a fixed point for $\Phi$. Let $\xi\in\Omega(\rho)$ and let $\zeta\in co\; \gamma(\xi)$, say $\zeta=\sum_{i=1}^n\lambda_i \Phi^{s_i}(\xi)$, $0\leq\lambda_i\leq 1$, $\sum_i\lambda_i=1$, $s_i\geq 1$, $i=1,\dots, n$. We have
$$\frac{1}{k}\sum_{m=0}^k \Phi^m(\zeta)=\sum_{i=1}^n \lambda_i \frac{1}{k}\sum_{m=s_i}^{k+s_i}\Phi^m(\xi)=$$
$$=\sum_{i=1}^n\lambda_i\frac{1}{k}\Big[ \sum_{m=0}^k \Phi^m(\xi) -\sum_{l=0}^{s_i}\Phi^l(\xi)+\sum_{m=k}^{k+s_i}\Phi^m(\xi)\Big]=$$
\begin{equation}
\frac{1}{k}\sum_{m=0}^k \Phi^m(\xi) +\sum_{i=1}^n\lambda_i\frac{1}{k}\Big[-\sum_{l=0}^{s_i}\Phi^l(\xi)+\sum_{m=k}^{k+s_i}\Phi^m(\xi)\Big]
\end{equation}
Hence since $\gamma(\xi)$ is bounded,
$$\limsup_{k\to\infty} d(\frac{1}{k} \sum_{m=0}^k \Phi^m(\xi),\eta_0)= \limsup_{k\to\infty} d(\frac{1}{k} \sum_{m=0}^k \Phi^m(\zeta),\eta_0)=$$
\begin{equation}
=\limsup_{k\to\infty} d(\frac{1}{k} \sum_{m=0}^k \Phi^m(\zeta),\Phi^m(\eta_0))\leq d(\zeta,\eta_0)
\end{equation}
But $\Omega(\rho)$ is minimal so we get that $\zeta\in \ov{co\; \gamma(\xi)}=\ov{co\;\Omega(\rho)}$ and $\eta_0\in\ov{co\; \Omega(\rho)}$. Therefore
\begin{equation}
\inf_{\zeta\in co\; \gamma(\xi)}d(\zeta,\eta_0)=0
\end{equation}
So we have $\eta_0=\lim_{k\to\infty} \frac{1}{k} \sum_{m=0}^k \Phi^m(\xi)$ for every $\xi\in\Omega(\rho)$ and therefore for every $\xi\in\ov{co\; \Omega(\rho)}$. In particular $\eta_0$ is the unique fixed point for $\Phi$ in $\ov{co\; \Omega(\rho)}$.

\qed

{\bf Proposition \ref{lvgaans}.} a) There exists a non-expansive projection $\tau$ onto $\Omega_\Phi$. b) We have that $\Omega_\Phi$ is convex and $\Phi$ restricted to such set is an isometry.

\bigskip

The proof uses the following lemma, which can be seen in \cite{lemmens}.

\begin{lemma}
Every $\Phi: D(\mathbb{C}^N)\to D(\mathbb{C}^N)$ CP map is such that every subsequence of $\{\Phi^k\}$ has a convergent subsequence that converges uniformly on compact subsets of $D(\mathbb{C}^N)$.
\end{lemma}

{\bf Proof of proposition \ref{lvgaans}.} a) By the above lemma there exists a subsequence $\{\Phi^{k_i}\}$ of $\{\Phi^k\}$ that converges uniformly on compact subsets of $D(\mathbb{C}^N)$. By passing to a subsequence and applying the lemma again we may assume that $k_{i_{j+1}}-k_{i_j}\to\infty$ as $j\to\infty$, and $\{\Phi^{k_{i_{j+1}}-k_{i_j}}\}$ converges uniformly on compact subsets of $D(\mathbb{C}^N)$. Define $\tau: D(\mathbb{C}^N)\to D(\mathbb{C}^N)$,
\begin{equation}
\tau(\rho):=\lim_{j\to\infty}\Phi^{k_{i_j+1}-k_{i_j}}(\rho),\;\;\; \rho\in D(\mathbb{C}^N)
\end{equation}
To see that $\tau$ is a projection onto $\Omega_\Phi$ first note that $\tau(\rho)\in \Omega_\Phi$ for every $\rho\in D(\mathbb{C}^M)$ so we have $\tau(D(\mathbb{C}^N))\subseteq\Omega_\Phi$. Let $\eta\in \Omega_\Phi$. It follows from proposition \ref{daf_prop1} that $\eta\in \Omega(\eta)$ and as $\gamma(\eta)$ has a compact closure, $\Phi(\Omega(\eta))=\Omega(\eta)$. Therefore there exists for each $k\geq 1$ a point $\eta_k\in\Omega(\eta)$ such that $\Phi^k(\eta_k)=\eta$. Denote by $g$ the pointwise limit of $\{\Phi^{k_i}\}$ and let $\{\eta_{l_j}\}$ be a convergent subsequence of $\{\eta_{k_i}\}$ with limit $\zeta\in\Omega(\eta)$ (such a subsequence exists as $\Omega(\eta)$ is compact). Then note that
\begin{equation}
d(\Phi^{l_j}(\eta_{l_j}),g(\zeta)) \leq d(\Phi^{l_j}(\eta_{l_j}),g(\eta_{l_j}))+ d(g(\eta_{l_j}),g(\zeta))\to 0,
\end{equation}
as $j\to\infty$, since $\{\Phi^{l_j}\}$ converges uniformly to $g$ on $\Omega(\eta)$. Therefore $g(\zeta)=\eta$. Now note that
\begin{equation}
\Phi^{k_{i_j+1}}(\zeta)=\Phi^{k_{i_j+1}-k_{i_j}}(\Phi^{k_{i_j}}(\zeta))
\end{equation}
for each $i\geq 1$. The LHS converges to $g(\zeta)$ whereas the RHS converges to $\tau(g(\zeta))$. Therefore $\tau(\eta)=\eta$ for all $\eta\in\Omega_\Phi$, so $\tau$ is a non-expansive projection onto $\Omega_\Phi$.

\bigskip

b) We have for all $\rho, \eta\in \Omega_\Phi$,
\begin{equation}
d(\rho,\eta)=d(\tau(\rho),\tau(\eta))\leq\lim_{j\to\infty}d(\Phi^{k_{i_j+1}-k_{i_j}}(\rho),\Phi^{k_{i_j+1}-k_{i_j}}(\eta))\leq d(\Phi(\rho),\Phi(\eta))\leq d(\rho,\eta)
\end{equation}
Let $\tau: D(\mathbb{C}^N)\to\Omega_\Phi$ be the projection given in a). For every $\rho,\eta\in \Omega_\Phi$ and $\leq\lambda\leq 1$ we have
\begin{equation}
\lambda\rho+(1-\lambda)\eta=\lambda\tau(\rho)+(1-\lambda)\tau(\eta)=\tau(\lambda\rho+(1-\lambda)\eta)\in\Omega_\Phi
\end{equation}

\qed

\end{document}